\documentstyle[prl,aps,preprint]{revtex}

\begin{document}
\draft
\title{An axially symmetric solution of metric-affine gravity} 
\bigskip

\author{E.J. Vlachynsky\footnote{Permanent address: Department of Mathematics, 
University of Newcastle, Newcastle, NSW 2308, Australia}, 
R. Tresguerres\footnote{Permanent address: Consejo Superior de 
Investigaciones Cientificas, Serrano 123, 28006 Madrid, Spain}, 
Yu.N. Obukhov\footnote{Permanent address: Department of Theoretical
Physics, Moscow State University, 117234 Moscow, Russia}, and F.W. Hehl}
\address{Institute for Theoretical Physics,
University of Cologne\\D-50923 K{\"o}ln, Germany}

\maketitle
\bigskip

\begin{abstract}

We present an exact stationary {\it axially symmetric} vacuum solution of
metric-affine gravity (MAG) which generalises the recently reported
spherically symmetric solution. Besides the metric, it carries
nonmetricity and torsion as post-Riemannian geometrical
structures. The parameters of the solution are interpreted as mass and
angular momentum and as dilation, shear and spin charges.
\end{abstract}
\bigskip\bigskip
\bigskip

The geometry of MAG (`metric-affine gravity') is described by 
the curvature two-form $R_{\alpha}
{}^{\beta}$, the nonmetricity one-form $Q_{\alpha\beta}$, and the torsion 
two-form $T^{\alpha}$ which are the gravitational field strengths for 
linear connection, metric and coframe, respectively. The corresponding 
physical sources are the canonical energy-momentum and hypermomentum 
three-forms. The latter includes the dilation, shear and spin currents
assosiated to matter. The field equations and the formalism are 
comprehensively described in \cite{PR}.

In the recent paper \cite{magexact}, we presented an exact spherically
symmetric solution of MAG which has a mass and three nontrivial 
hypermomentum charges (dilation, shear and spin). Here we
generalise this result and describe an axially symmetric solution. 

We start from the Lagrangian \cite{magexact},
\begin{eqnarray}
V&=& \frac{1}{2\kappa}\,\left[-a_0\,R^{\alpha\beta}
\wedge\eta_{\alpha\beta}-2\lambda\,\eta+ T^\alpha\wedge{}^*\!
\left(\sum_{I=1}^{3}a_{I}\,^{(I)}T_\alpha\right)\right.
\nonumber\\&+&\left.
2\left(\sum_{I=2}^{4}c_{I}\,^{(I)}Q_{\alpha\beta}\right)\wedge
\vartheta^\alpha\wedge{}^*\!\, T^\beta + Q_{\alpha\beta}\wedge{}^*\!
\left(\sum_{I=1}^{4}b_{I}\,^{(I)}Q^{\alpha\beta}\right)\right]
\nonumber\\&-& \frac{z_{4}}{2}\,R^{\alpha\beta} 
\wedge{}^*\!\,^{(4)}Z_{\alpha\beta}\,,\label{lagr}
\end{eqnarray}
where $a_0,...,a_3, b_1,...,b_4, c_2, c_3, c_4, z_4$ are coupling
constants. We choose the cosmological constant $\lambda$ to be positive
for an overall repulsion. 
As usual, the star $^*$ denotes the Hodge dual, and $\kappa$ is the
Einstein gravitational constant. The metric signature is $(-+++)$.
The fourth irreducible piece of the curvature reads $^{(4)}Z_{\alpha\beta}
=R_{\gamma}{}^{\gamma}g_{\alpha\beta}/4$ (see \cite{PR} for details). 
Particular cases of (\ref{lagr}) were considered earlier in 
\cite{hls,pono,tres1,tres2,tw}. 

Let us choose standard Boyer-Lindquist coordinates $(t,r,\theta,\phi)$. 
With the local Min\-kowski metric, $o_{\alpha\beta}=
\hbox{diag}(-1,1,1,1)$, we take as an ansatz the standard coframe for 
the Kerr solution with cosmological constant:
\begin{eqnarray}
\vartheta ^{\hat{0}} &=&\,\sqrt{{\Delta}\over{\Sigma}}\,
\left(\,d\, t  -j_0\sin ^2\theta\,d\phi\,\right),\nonumber\\
\vartheta ^{\hat{1}} &=&\,\sqrt{{\Sigma}\over{\Delta}}\;d\,r,\nonumber\\
\vartheta ^{\hat{2}} &=&\,\sqrt{{\Sigma}\over f}\;d\,\theta,\nonumber\\ 
\vartheta ^{\hat{3}} &=&\,\sqrt{f\over{\Sigma}}\,\sin\theta\,
\left[\,- j_0 d\, t  +\left(\,r^2+j_0^2\,\right) d\phi\,\right],\label{frame}
\end{eqnarray}
where $\Delta=\Delta(r), \Sigma=\Sigma(r,\theta), f=f(\theta)$, and 
$j_0$ is a constant.

Similarly to \cite{magexact}, we assume that nonmetricity and torsion
are represented by three one-forms (the Weyl covector, the
third irreducible nonmetricity piece and the torsion trace) so that
\begin{eqnarray}
Q_{\alpha\beta}&=&\,u(r,\theta)\;o_{\alpha\beta}\,\vartheta^{\hat{0}} + 
\frac{1}{9}\,v(r,\theta)\,\left(4\vartheta_{(\alpha}e_{\beta )}\rfloor
\vartheta^{\hat{0}}-o_{\alpha\beta}\vartheta^{\hat{0}}\right),
\label{nonmet}\\T^\alpha&=&\frac{1}{3}\,\tau(r,\theta)\;
\vartheta^\alpha\wedge\vartheta^{\hat{0}}.\label{tor}
\end{eqnarray}

Substituting the ansatz (\ref{frame})-(\ref{tor}) into the field equations
for the Lagrangian (\ref{lagr}) (we refer the readers to \cite{PR} for the
general form of the MAG field equations, (5.5.4)-(5.5.5), and to 
\cite{magexact} for the detailed discussion of the model (\ref{lagr})) one 
finds:
\begin{equation}
u=k_0N{r\over{\sqrt{\Delta\Sigma}}},\quad
v=k_1N{r\over{\sqrt{\Delta\Sigma}}},\quad
\tau=k_2N{r\over{\sqrt{\Delta\Sigma}}},\label{sol1}
\end{equation}
\begin{eqnarray}
\Delta &=&r^2+j_0^2-2\kappa Mr-{{\lambda}\over 3a_0}\,
r^2\left(r^2+j_0^2\right) +z_4{\kappa (k_0N)^2\over 2a_0},\label{sol2a}\\ 
\Sigma &=&\,r^2 +j_0^2\cos^2\theta,\label{sol2b}\\ 
f&=&\,1+{{\lambda}\over 3a_0}\,j_0^2\cos ^2\theta\,.\label{sol2c}
\end{eqnarray}
Here $M$ and $N$ are the integration constants which describe, respectively, 
the mass and the nonmetricity-torsion charges of the source. The 
coefficients $k_0,\,k_1,\,k_2$ are constructed in terms of the 
coupling constants,
\begin{eqnarray}
k_0 &=& \left({a_2\over 2}-a_0\right)(8b_3 + a_0) - 3(c_3 + a_0 )^2\,,
\label{k0}\\
k_1 &=& -9\left[ a_0\left({a_2\over 2} - a_0\right) + 
(c_3 + a_0 )(c_4 + a_0 )\right]\,,
\label{k1}\\
k_2 &=& {3\over 2} \left[ 3a_0 (c_3 + a_0 ) + 
(8b_3 + a_0)(c_4 + a_0 )\right]\,,\label{k2}
\end{eqnarray}
and, as in \cite{magexact}, we find
\begin{equation}
b_4=\frac{a_0k+2c_4k_2}{8k_0}\,,\qquad\hbox{with}\qquad 
k:=3k_0-k_1+2k_2\,.\label{b4}
\end{equation}

The physical interpretation of the parameters of the solution as
described above is clear:
$M$ and $j_0$ represent the mass and the angular momentum of Kerr type,
while $(k_0N), (k_1N)$, and $(k_2N)$ describe, respectively, the dilation 
(`Weyl') charge of the fourth irreducible nonmetricity piece, the shear 
charge of the third irreducible nonmetricity piece and the spin charge 
of the second irreducible torsion part. 

Work is in progress on the generalisation of our solution to the MAG
version of the Pleba\'nski-Demia\'nski family \cite{PD}.

\bigskip
\noindent{\bf Acknowledgments}
This research was supported by the Deutsche Forschungsgemeinschaft (Bonn)
under project He-528/17-1 and by the Graduate College ``Scientific 
Computing'' (Cologne-St.Augustin).

\end{document}